\documentclass[twocolumn]{aa}
\usepackage{graphics}
\usepackage{epsfig}
\usepackage{latexsym}
\usepackage{natbib}

\def\E$\gamma${E_\$\gamma$}

\def\deg       {$^{\circ}$}

\def \gray     {$\gamma$-ray}
\def \grays    {$\gamma$-rays}

\begin{document}

%

\title{Spectral constraints on unidentified EGRET gamma-ray sources
 from COMPTEL MeV observations}

\author{ S.~Zhang\inst{1,2},
        W.~Collmar\inst{2},
        W.~Hermsen\inst{3,4},
         V.~Sch\"onfelder\inst{2}
}
 
\institute{ Laboratory of Cosmic Ray and High Energy Astrophysics, 
              Institute of High Energy Physics,
             P.O. Box 918-3, Beijing 100039, China
             \and
Max-Planck-Institut f\"ur extraterrestrische Physik,
               P.O. Box 1603, D-85740 Garching, Germany 
            \and
             SRON National Institute for Space Research, Sorbonnelaan 2, 
              NL-3584 CA Utrecht, The Netherlands
            \and
Astronomical Institute 'Anton Pannekoek', University of Amsterdam,
Kruislaan 403, 1098 SJ Amsterdam, The Netherlands
          }

\offprints{W.~Collmar}
\mail{wec@mpe.mpg.de}

\date{Received 13 November 2003 / Accepted 31 March 2004}

\abstract{
We investigated the MeV properties of 173 unidentified or
only tentatively identified EGRET sources listed in the third
EGRET catalogue, by analyzing the simultaneously collected
COMPTEL MeV data for each individual source. 
The sources can generally be divided into 4 groups. In this paper
we focus on one of these, a group of 22 EGRET sources for 
which we can provide additional constraining information:
their spectral extrapolations from the energy range above 100 MeV
towards lower energies overshoot the fluxes or upper limits derived 
simultaneously at MeV energies. 
This means that for these sources a spectral turnover/break 
between 1 MeV and 100 MeV  is required. 
At least two of these sources, but most likely 
the majority of this sample, have the maxima of their
gamma-ray luminosities in this energy band. 
The sources have rather soft EGRET spectra
 (average photon index $=$ 2.72$^{+0.08}_{-0.11}$),
and seem to spatially cluster in the inner Galaxy. 
Variability analyses revealed 11 out of the 22 sources to be
significantly variable. 
Object classes proposed as possible counterparts for the unidentified 
EGRET sources are discussed in the light of these additional constraints.   
 
}

\titlerunning{Spectral constraints on unidentified EGRET ...}
\authorrunning{S.~Zhang et al.}
\maketitle

\section{Introduction}

One of the biggest mysteries left by the Compton Gamma-Ray Observatory
(CGRO, 1991-2000) is that a large number of
$\gamma$-ray sources detected by the different CGRO experiments,
in particular EGRET, still remain unidentfied.
 The EGRET experiment measured \grays\ above 30~MeV, most 
sensitively above 100~MeV.
Out of the 271 sources listed in the third 
EGRET catalogue \citep{Hartman99}, 170 are unidentified and 27
are only tentatively identified. 
Several classes of objects have been proposed as possible 
counterparts for those unidentified EGRET sources. 
Sources located at high galactic latitudes and being
time variable are believed to be active galactic nuclei (AGN),
in particular blazars.
Sources, located at lower galactic latitudes, being steady and having 
low \gray\ fluxes, are found to coincide spatially with  
the Gould Belt \citep{Gehrels00}. Some other low-latitude sources 
show positional correlations with supernova remnants (SNRs)
and OB associations \citep[e.g.][]{Romero99}. 
Steady sources with hard \gray\ spectra seem to be good candidates 
for young $\gamma$-ray pulsars with ages of less than
10$^{6}$ years \citep{Zhang00a}. 
Several sources (mainly located  $|$b$|$ $<$ 10\deg) might indicate
a new class of $\gamma$-ray emitting objects \citep{Torres01},
because they do not coincide with any potential counterpart objects.

The COMPTEL experiment aboard CGRO is sensitive to \gray\  photons 
between 0.75 and 30~MeV, thereby covering the softer \gray\ band
adjacent to the EGRET one.
Apart from transient $\gamma$-ray bursts, 
unidentified \gray\ sources and AGN are the majority of the COMPTEL
source detections.
The first COMPTEL catalogue \citep{Schonfelder00} lists
 10 AGN and 9 unidentified
$\gamma$-ray sources; the sum of the rest (radio pulsars,
 stellar black-hole candidates,
SNRs, and $\gamma$-ray line sources) is about 12. 
Since COMPTEL and EGRET were mounted parallel on CGRO and both had a 
large field of view (the COMPTEL one being larger than the EGRET one),
COMPTEL and EGRET observed simultaneously the sa\-me sky region.

To gain further knowledge on the unidentified 
EGRET sources, and to probe their nature,
we analyzed the contemporaneous
COMPTEL data on the unidentified EGRET sources to
supplement the EGRET results.
In this paper, we report the discovery of 
a subgroup of the unidentified EGRET sources
whose \gray\ spectra are constrained by the MeV data:  
their spectral energy distributions have at least an MeV break
but most likely an MeV peak.
The paper is organized as follows: in Section 2 we briefly describe
the COMPTEL instrument, the applied data analysis methods and
the observational concept of CGRO, in Section 3 we present the analysis
results and discuss them in Section 4. In Section 5, we finally 
present the conclusions.

\section{Instrument, Data Analysis and CGRO Observations}
The Compton telescope COMPTEL (0.75-30~MeV)
had an energy-dependent energy and angular resolution of 5\% -- 8\% (FWHM)
and 1.7\deg -- 4.4\deg (FWHM), respectively. Its field of view is
circular and covers $\sim$1 steradian.
Imaging in its large field of view is possible with a location 
accuracy (flux dependent) of the order of 1\deg -- 3\deg. 
For details on the experiment see \citet{Schonfelder93}.

Skymaps and source parameters, like detection significances,
fluxes, and flux errors, can be obtained via the maximum likelihood
method, which is implemented in the standard COMPTEL data analysis package.
The detection significance is derived from the 
quantity -2ln$\lambda$, where $\lambda$ is the ratio of the likelihood L$_{0}$
(background) to the likelihood L$_{1}$ (source + background). 
The quantity -2ln$\lambda$ has a $\chi_{1}^{2}$ distribution, 
if only the flux at a given source position is estimated \citep{Boer92}. 
The detection significance can be conservatively calculated by the ratio
of flux to flux error. This approach is adopted in this paper for estimating
the source detection significances. 

In fitting the fluxes of the relevant  EGRET sources, 
nearby prominent COMPTEL sources are taken into account
 by fitting simultaneously their fluxes. 
An estimate for the instrumental background of COMPTEL is derived by using
the standard filter technique in the COMPTEL data space \citep{Bloemen94}. 
The celestial background components, galactic
and extra-galactic diffuse $\gamma$-ray
radiation, also have been taken into account by model fitting. 
In the presented analyses we applied instrumental point spread functions 
assuming an E$^{-2}$ power-law shape for the source spectra.
 
CGRO observations were organized in so-called 'Mission Phases'
and 'Viewing Periods (VPs)'. A 'Mission Phase' covers typically 1 year of data 
and contains many VPs, which typically last for 1 to 2 weeks each. 
For each EGRET source of interest we analyzed the simultaneously 
collected COMPTEL data. 
To allow combining of COMPTEL and EGRET results, we analyzed
the MeV data for periods for which the EGRET spectral index was 
estimated. 
Many EGRET source results are published in the sum of the CGRO
Mission Phases 1 to 4, noted as 'P1234' in the third EGRET catalogue.
For this time period we generated COMPTEL all-sky data, 
and derived the flux results by fitting sources at the
relevant positions. As noted above, the fit was performed
by including 1) models for the cosmic diffuse radiations,  2) the 
strongest MeV source (the Crab), and 3) neighboring COMPTEL sources
on a case by case basis. 

To compare our COMPTEL source fluxes with the EGRET ones,
we plotted the best-fit power-law shapes to the EGRET source spectra 
with 1$\sigma$ errors and extrapolated these below 100~MeV into 
the COMPTEL band. 
Systematic errors of 10\% are included in these EGRET spectra.
The COMPTEL flux values are given for 4 standard 
COMPTEL bands (0.75-1, 1-3, 3-10, 10-30 MeV).  For a detection
significance $<$2$\sigma$ 
the source flux is presented by a 2$\sigma$ upper limit, otherwise 
by a flux point. Additionally, when comparing with the EGRET spectral
extrapolations, the COMPTEL error bars and subsequently also the
upper limits are enlarged by 20 percent to account for 
systematic errors.

\section{Results}
\subsection{General Results}
We investigated all unidentified or tentatively identified EGRET sources 
for which spectral indices are
given in the third EGRET catalogue. By excluding 5 artifacts near the
Vela pulsar and 1 artifact near Crab \citep{Thompson01}, 
an ensemble of 173 sources was selected for our analyses.
In some sky regions the  unidentified EGRET sources are 'crowded'.
For cases where they are closer than the COMPTEL location accuracy,
no precise and unambiguous MeV flux could be attributed to individual 
EGRET sources. In these cases only one MeV source was fitted and 
its flux value was treated as upper limit for the different EGRET sources.
 
\begin{figure}[bt]
\epsfig{figure=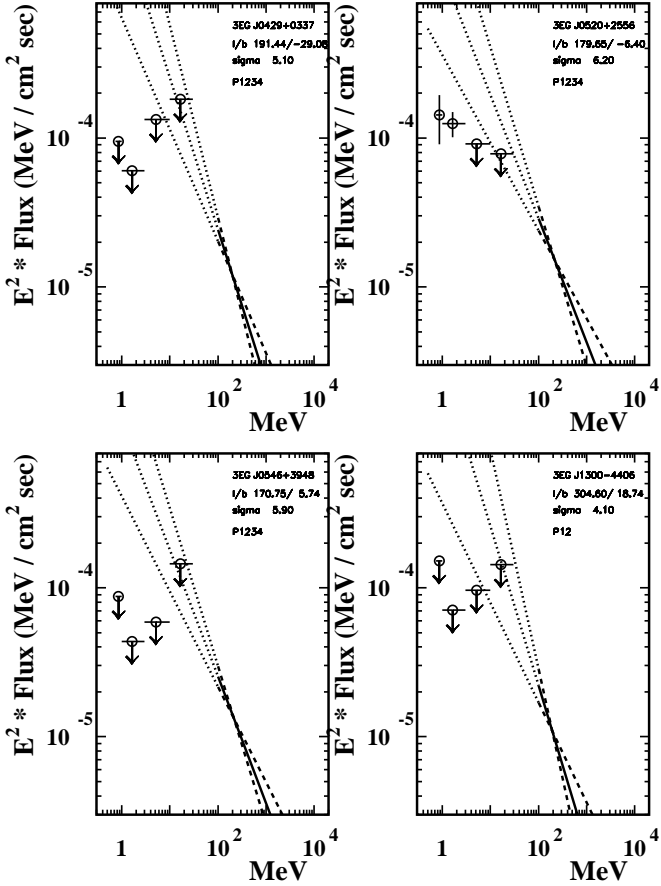,width=8.8cm,clip=}
\caption{
Combined simultaneous COMPTEL/EGRET energy spectra of 4 
sources out of our sample of 22 sources (see Table~1).
The inserted text gives the EGRET 3rd catalogue informations
(Hartman et al. 1999) on the source
(source name, sky location, EGRET detection significance, 
and detection period).
The solid line represents the best-fit EGRET spectrum 
above 100~MeV,
the dashed lines its 1$\sigma$ error in spectral index,
and the dotted lines the spectral extrapolations below 
100~MeV down to 0.75~MeV. 
The required spectral changes at MeV energies are obvious.
The COMPTEL upper limits are 2$\sigma$ and the error bars
on the flux points are 1$\sigma$.
}
\label{fig:spectra}
\end{figure}

In most cases, we derive only upper flux limits or marg\-inal 
($<$4$\sigma$) hints for MeV emission.
We find significant MeV detections in 4 sky regions, 
three of them are locations of already known 
unidentified MeV sources ((l,b)= (358.5, 0.5), (311.5, -2.5), (18.5, -0.5)). 
The fourth one, at $\sim$(l,b)=(188.7, -4.4) and dubbed GRO~J0550+19
in Bronsveld et al. (2002), is near the location of the Crab, 
which is by far 
the strongest MeV source. Detailed studies of these sky regions have been 
reported earlier or are in progress \citep{Strong01, Zhang02, Bronsveld02, Collmar04}.

In general our results allow the sources to be divided into 4 groups.
For the majority of sources ($\sim$120) COMPTEL cannot provide any constraints
on the EGRET spectra (group 1). We detect evidence 
(detection significance $>$2$\sigma$) for about 20
sources whose MeV fluxes 
are consistent with the EGRET extrapolations, showing that the 
measured EGRET spectra extend into the MeV band (group 2). 
A third group contains a few sources indicating a spectral upturn 
of the EGRET spectra at MeV energies, suggesting the presence of an
additional spectral component. Finally, we found 22 sources whose COMPTEL fluxes or
flux upper limits are below the expected fluxes based on extrapolations of 
their EGRET spectra, 
requiring a spectral turnover/\-break at MeV energies (group 4).      
In this paper we want to concentrate on the sources of group 4
for which COMPTEL can provide meaningful constraints on the EGRET spectra.
Details on the three other groups will be given in a later paper.

\subsection{Sources with spectral constraints in the MeV band}
  
Applying the method described above, we found a subgroup of 22 
unidentified (no tentatively identified source belongs to our sample)
 EGRET sources
for which a spectral flattening at MeV energies is required.
Fig.~1 shows four typical examples.
All 22 sources are listed in Table~1
including the relevant source parameters. 
At least two sources, 3EG~J1638-5515 and 3EG~J1823-1314, have a $\gamma$-ray
luminosity peak in the 1-100 MeV band.
Their luminosity or luminosity upper limit in the COMPTEL 
1-3 MeV band is significantly lower than measured luminosity 
values at energies between 10 and 100 MeV, i.e. the COMPTEL 
spectral shape is harder than E$^{-2}$, while the EGRET one is softer 
than E$^{-2}$. In fact, most
likely the majority of sources in this sample has its maximum luminosity
in this range, since we know that generally no strong hard X-ray sources
have been found in EGRET error boxes.

\begin{figure*}[t]
\centering
\epsfig{figure=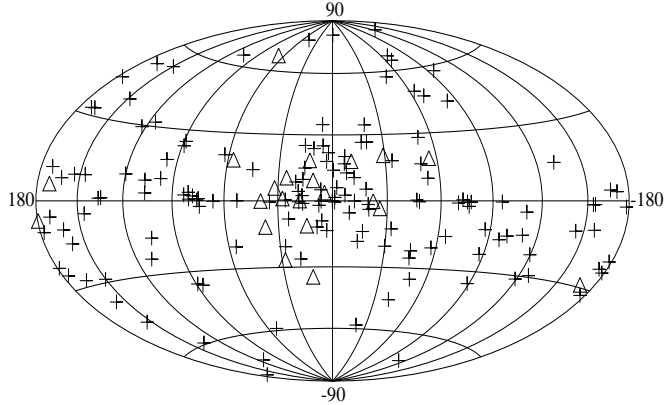,width=8.8cm,height=7.0cm,clip=}
\caption{
The sky distribution of 173 unidentified and tentatively identified 
EGRET sources ($+$, $\bigtriangleup$) in galactic coordinates.
The triangles ($\bigtriangleup$)
represent the 22 EGRET sources of our sample.
A clustering towards the inner galaxy is indicated.
Especially, 20 of the 22 sources are located at 
$|$b$|$ $<$ 30$^{\circ}$, suggesting a galactic origin.
} 
\label{fig:map}
\end{figure*}

\begin{table*}[t]
\caption{
Catalogue of the 22 unidentified 
EGRET sources which -- by inclusion of the COMPTEL results -- 
have to have a spectral break or their \gray\ luminosity maxima at MeV energies.
The source parameters of the first six columns are taken
from the third EGRET catalogue, column 7 and 8 are for  
the variability analysis, and the last 4 columns give the COMPTEL fluxes/upper limits
in the 4 standard energy bands. The abbreviations are the following:
\newline{\it{Name}} -- 3EG source name.
\newline{\it{l. b.}} -- source coordinate of galactic longitude and latitude.
\newline{\it{F}} -- EGRET source flux and error in units of 10$^{-8}$ ph cm$^{-2}$ s$^{-1}$. 
\newline{\it{$\alpha$}} -- photon spectral index and its  error.
\newline{\it{VP}} -- CGRO observation periods for which the spectral index is derived.
P1 means phase 1, P12 the combination of P1 and P2, and so on.
Observation 330+ is the combined VPs 330 and 332.
\newline{\it{$\tau$($\tau_{min}$,$\tau_{max}$)}}
 --  taken from \citet{Tompkins99} for measuring the fractional
variability of EGRET sources. $\tau_{\rm{min}}$ and $\tau_{\rm{max}}$ represent separately the 68$\%$ 
lower and upper limits on $\tau$.
\newline{\it{I}} -- index defined by \citet{Torres01} for measuring source 
variability with respect to $\gamma$-ray pulsars; derived by our
variability analysis on the published EGRET fluxes.
\newline{\it{Flux (10$^{-5}$ ph cm$^{-2}$ s$^{-1}$)}} -- COMPTEL fluxes and 
the flux unit in the four standard energy bands: 0.75-1, 1-3, 3-10 and 
10-30 MeV. The upper limits are 2$\sigma$ and the error bars are 1$\sigma$.
}
\begin{flushleft}
\begin{tabular}{cccccccccccc}\hline
\multicolumn{1}{c}{Name}&\multicolumn{1}{c}{l}&\multicolumn{1}{c}{b}&\multicolumn{1}{c}{F}&\multicolumn{1}{c}{$\alpha$}&\multicolumn{1}{c}{VP}&\multicolumn{1}{c}{$\tau$($\tau_{\rm{min}}$,$\tau_{\rm{max}}$)}&\multicolumn{1}{c}{I}&\multicolumn{4}{c}{Flux
(10$^{-5}$ ph cm$^{-2}$ s$^{-1}$)}\\  
\multicolumn{1}{c}{3EG
J}&\multicolumn{1}{c}{deg}&\multicolumn{1}{c}{deg}&\multicolumn{1}{c}{}&\multicolumn{1}{c}{}&\multicolumn{1}{c}{}&\multicolumn{1}{c}{}&\multicolumn{1}{c}{}&\multicolumn{1}{c}{0.75-1
}&\multicolumn{1}{c}{1-3 }&\multicolumn{1}{c}{3-10}&\multicolumn{1}{c}{10-30}\\  \hline
0429+0337 & 191.44  & -29.08 & 12.0$\pm$2.7 & 3.02$\pm$0.27  & P1234  & 0.00(0.00,0.45)  &1.6  & $<$3.2 &$<$4.0  &$<$3.1  &$<$1.2 \\
0520+2556 & 179.65  & -6.40  & 15.7$\pm$2.7 & 2.83$\pm$0.24  & P1234  & 0.00(0.00,0.31)  & 1.1 & 4.8$\pm$1.7 &8.3$\pm$1.6  &$<$2.1 
&$<$0.5 \\
0546+3948 & 170.75  & 5.74   & 13.7$\pm$2.6 & 2.85$\pm$0.21  & P1234  & 0.11(0.00,0.47)  & 1.7 & $<$2.9 & $<$2.9 & $<$1.4 &$<$0.9 \\
1300-4406 & 304.6   & 18.74  & 10.6$\pm$2.9 & 3.07$\pm$0.40  & P12    & 0.48(0.00,1.57)  & 3.0 & $<$5.0 &$<$4.7  & $<$2.2 &$<$1.0 \\
1424+3734 & 66.82   & 67.76  & 16.3$\pm$4.9 & 3.25$\pm$0.46  & P1     & 0.01(0.00,$\infty$)&1.9  & $<$6.4 &8.5$\pm$3.1  & $<$5.6 &$<$2.2 \\
1500-3509 & 330.91  & 20.45  & 10.9$\pm$2.8 & 2.99$\pm$0.37  & P1234  & 0.00(0.00,0.61)  &1.5  & $<$5.1 &$<$3.3  &$<$1.6  &$<$1.0 \\
1612-2618 & 349.40  & 17.90  & 92.2$\pm$27.7 & 2.71$\pm$0.23  & 423.   &1.78(0.76,11.74) &4.1  & $<$41.0 &$<$19.1 &$<$9.3  &$<$4.2 \\
1638-5515 & 334.05  & -3.34  & 67.3$\pm$14.2 & 2.56$\pm$0.21  & P2     &0.00(0.00,0.69)  & 2.4 & $<$15.6 &$<$7.0  &$<$6.7  &2.8$\pm$1.0 \\
1639-4702 & 337.75  & -0.15  & 53.2$\pm$8.7 & 2.50$\pm$0.18  & P1234  & 0.00(0.00,0.38)  &2.0  & $<$3.1 & $<$4.5 & $<$3.0 &$<$0.9 \\
1709-0828 & 12.86   & 18.25  & 12.6$\pm$3.2 & 3.00$\pm$0.35  & P1234  & 0.84(0.11,2.21)  &2.7  & $<$3.2 & $<$3.2 & $<$2.8 &$<$0.6 \\
1735-1500 & 10.73   & 9.22   & 196.3$\pm$48.8 & 3.24$\pm$0.47  & 231.0  & 1.09(0.00,10.14)&8.9 & $<$29.0 &$<$23.1  &$<$11.8  &$<$4.5 \\
1741-2312 & 4.42    & 3.76   & 33.1$\pm$5.9 & 2.49$\pm$0.14  & P12    & 0.52(0.18,1.03)  & 2.2 & $<$11.9 &$<$4.4  &$<$3.0  &$<$0.9 \\
1800-0146 & 25.49   & 10.39  & 26.1$\pm$6.1 & 2.79$\pm$0.22  & P34    & 0.00(0.00,0.48)  & 1.9 & $<$8.1 &$<$7.0  &$<$2.2  & $<$1.4\\
1823-1314 & 17.94   & 0.14   & 102.6$\pm$12.5& 2.69$\pm$0.19  & P3   & 0.72(0.40,1.37)   &3.0  & 10.2$\pm$2.9&5.6$\pm$2.8  & 3.7$\pm$1.2 &2.7$\pm$0.5 \\
1825+2854 & 56.79  & 18.03  & 34.3$\pm$10.9 & 4.47$\pm$1.15  & 9.2    & 73.19(2.59,$\infty$)& 2.5 & $<$30.1 &$<$13.9  & $<$8.6 &$<$2.2 \\
1828+0142 & 31.90   & 5.78   & 132.2$\pm$24.0 & 2.76$\pm$0.39  & 13.1   &3982.15(6.92,$\infty$)& 5.3  & $<$16.4 &$<$15.8  &$<$11.3  &$<$4.4 \\
1837-0423 & 27.44   & 1.06   & 310.4$\pm$63.7& 2.71$\pm$0.44  & 423.0  &12.01(2.17,$\infty$)&8.4   & $<$23.4 & $<$18.5 & $<$14.3 &$<$3.0 \\
1858-2137 & 14.21   & -11.15 & 11.2$\pm$2.6 & 3.45$\pm$0.38  & P1234  & 0.00(0.00,0.56)&2.8  & $<$4.2 & $<$5.0 & 1.6$\pm$0.7 &$<$0.6 \\
1903+0550 & 39.52   & -0.05  & 62.1$\pm$8.9 & 2.38$\pm$0.17  & P1234  &0.35(0.18,0.60)& 2.3  & $<$3.3 & $<$3.4 &$<$2.7  &$<$0.7 \\
1940-0121 & 37.41   & -11.62 & 41.0$\pm$10.7 & 3.15$\pm$0.39  & 330.+  &4.58(1.13,$\infty$)&4.0   & $<$16.6 &$<$10.9  &$<$6.8  &$<$1.8 \\
2020-1545 & 28.09   & -26.62 & 11.8$\pm$3.4 & 3.40$\pm$0.55  & P1     & 0.00(0.00,0.80)&0.9  & $<$5.2 &$<$7.7  & $<$2.6 &$<$1.0 \\
2034-3110 & 12.25   & -34.64 & 17.4$\pm$5.2 & 3.43$\pm$0.78  & P1     & 2.88(0.89,154.84)& 5.7 & $<$5.4 &$<$5.4  & $<$3.2 &$<$1.7 \\ \hline
\end{tabular}\end{flushleft}
\label{tab:flux}
\end{table*}

The spatial distribution of the 22 sources is shown in Fig.~2. 
A concentration at low galactic latitudes of 
$|$b$|$ $<$ 30$^{\circ}$ is apparent. The sources tend to  
concentrate in the inner galactic region. Such a distribution  
suggests galactic origins for most of these sources.

Fig.~3 shows correlation plots of spectral index versus flux at energies 
above 100 MeV for a) the 22 selected sources and b) the rest 
 ($\sim$151 sources) of the unidentified and tentatively identified EGRET sources.
No obvious correlation between spectral 
index and flux is visible for both source groups. 
The linear fit results in average photon indices of 2.72$^{+0.08}_{-0.11}$
for the 'break' sample and 2.13 $\pm$ 0.03 for the rest.
The softer energy spectra for the 'break' sample might be a selection
effect due to the worse detection sensitivity of COMPTEL compared to EGRET. 
For sources having similar flux levels in the EGRET band, those with a softer
energy spectrum are more likely to be constrained by the COMPTEL data,
because the extrapolation of the EGRET spectrum into the COMPTEL band will reach  
higher flux values. For the harder sources, the spectral extrapolation might 
go below the COMPTEL sensitivity limits making COMPTEL constraints impossible.

\subsection{Variability}
There are  three approaches to estimate the flux 
variability of $\gamma$-ray sources. The so-called V method \citep{McLaughlin96} is 
based on the $\chi^2$-test. It is affected by the source detection significance,
which is in \gray\ analyses/data often low.
For a low-significance source with intrinsic variability a small V value is derived.
On the other hand, a large V value can either be due to real source variability 
or due to large systematic effects.
To overcome these systematic problems, \citet{Zhang00a} 
and \citet{Torres01} suggested to use the so-called I-index instead. 
It is defined as the ratio of the measured source variability 
to those of $\gamma$-ray pulsars, I=$\mu_{\rm{source}}$/$\mu_{\rm{pulsars}}$, 
which are considered to be intrinsically constant. $\mu$ is the ratio
of the standard deviation of the measured flux values to the weighted
mean flux. We regard a source as variable,
if I-index $>$ 2.5 (corresponding to a 3$\sigma$ significance
for variability), a source as stable, if I-index $<$ 1.5, and a source 
as dubious, if its I-index value is in between.
We calculated the I-indices for the 22 sources of our sample
according to  \citet{Torres01} 
by using the flux values given in the 3rd EGRET source catalogue.
The results are listed in Table~1. We found 10 sources to 
have an I-index $>$ 2.5 (i.e. variable) and two to have an I-index
$<$ 1.5 (i.e. constant).  

\citet{Tompkins99} investigated the source variability via the $\tau$-method, 
which is defined as $\tau$=$\sigma$/$\mu$, 
where $\sigma$ is the standard deviation of the fluxes and $\mu$ the
average value. The idea is to overcome systematic effects by comparing to the 
$\tau$ distribution for sources of known nature. 
The source fluxes are assumed to have a Gaussian distribution with 
the parameters $\tau$$\times$$\mu$ (width) and $\mu$ (mean value).
These parameters are estimated by the maximum likelihood method.
By  comparing to the $\tau$ distribution of blazars, a source is
regarded variable if its 68$\%$ lower limit is greater than 1.
This results in four of our sources being variable. 
Three of them  have an I value $>$ 2.5 and one, 3EG J1825+2854,
has an I value of 2.49. These results show that the I- and
$\tau$-methods provide consistent results. 
All other 18 sources, with $\tau_{\rm{max}}$ values larger than 0.3 and 
$\tau_{\rm{min}}$ values less than 1, 
would be classified as dubious, according to the definition by \citet{Tompkins99}.
Nolan et al. (2003) adopted the same $\tau$-method, only applying slight 
changes/modifications, to investigate the time variability of the 
EGRET $\gamma$-ray sources. 
Their approach resulted in some numerical differences to Tompkins. 
According to the classification by Nolan et al. (2003), four sources of our Table 1,   
3EG J1612-2618,1828+0142, 1837-0423, 1940-0121 and 2034-3110, are 
variable. These sources have I-indices $>$ 2.5. 
All others, with the exception of 
3EG J1424+3734, which is not included in Nolan et al. (2003) due to 
their selection criteria on the observations, are classified as dubious.
Since the $\tau$ values are estimated by lumping the VPs within one month
while the I-index method  uses individual VPs, typically ranging 
from several days to two weeks, the two methods may estimate the source
variability on different time scales.
We use the I-index to classify the source variability.
This yields, as mentioned above, 10 variable sources. Because of the large 
I-index value (2.49) and the variability-indicating  $\tau$ value ($\tau_{\rm{min}}$ $\sim$2.59), 
we consider also 3EG J1825+2854 as a variable source.
The two sources with I-index $<$ 1.5 are considered dubious
due to their $\tau$ values. Thus we have a sample of 11 variable sources 
and of 11 dubious sources. 

Fig.~4 shows correlation plots of spectral photon index versus I-index 
for the 22 sources of our sample and the remaining 151 sources.
For the sample of the 151 sources, the linear correlation coefficient is
derived to be $\sim$0.15, suggesting  that the softer sources tend to be
more variable. This is consistent with Torres et al. (2001), who found 
that the most-variable unidentified EGRET sources near the galactic plane tend 
to have steep spectra. The 22 sources do not show such a trend.
A linear correlation coefficient of $\sim$0.03 indicates an uncorrelated sample.
This might either discriminate this sample of 22 sources from the rest, or is
due to the relatively poor statistics.      

\begin{figure}[t]
\centering
\epsfig{figure=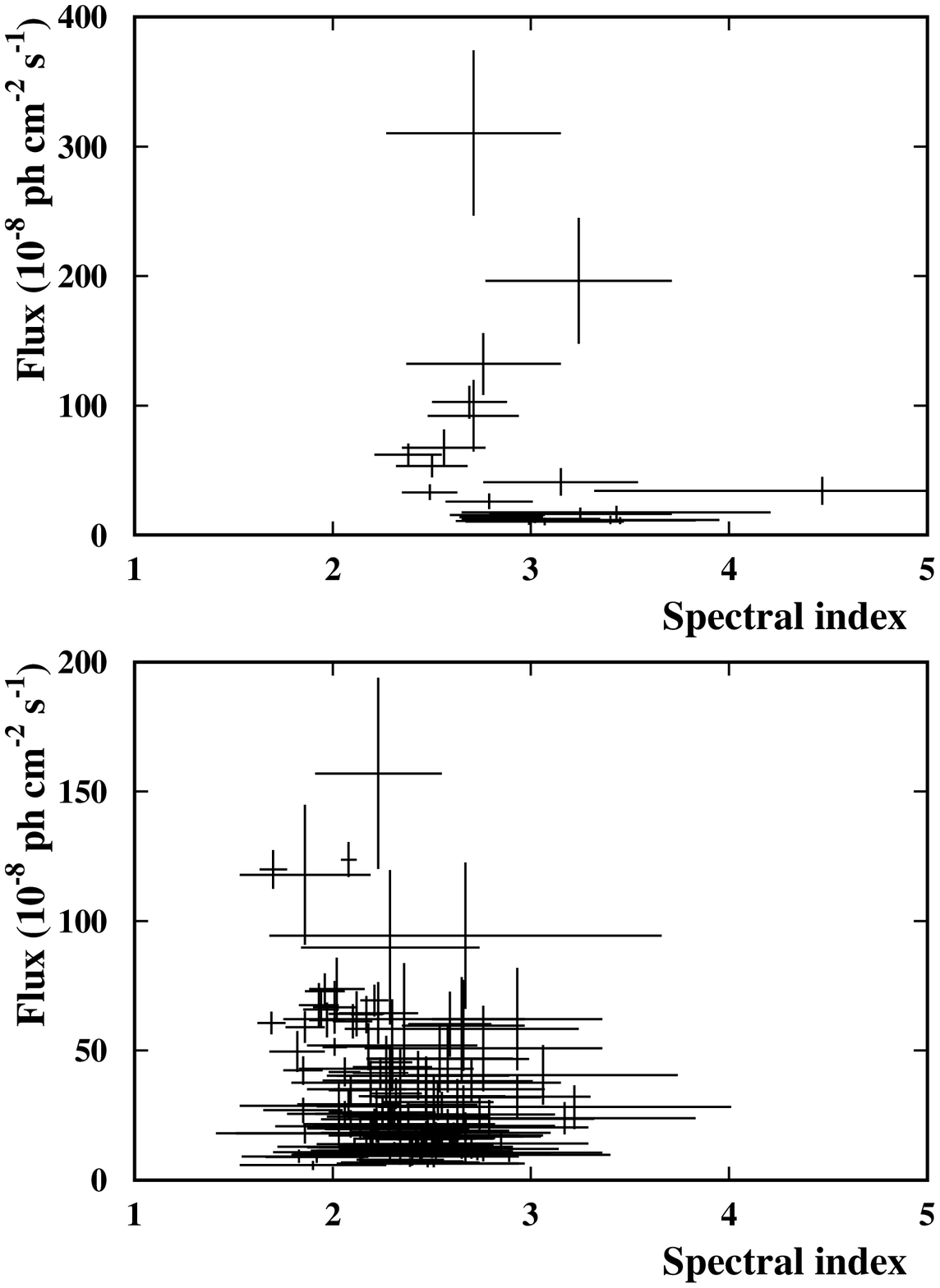,width=8.8cm,clip=}
\caption{
Correlation plots of spectral photon index versus flux (EGRET results)
for our sample of 22 sources (upper panel) and the rest of 
151 unidentified and tentatively identified EGRET sources (lower panel).
The sources of our sample typically have softer spectra.
} 
\label{fig:spec_flux}
\end{figure}

\begin{figure}[t]
\centering
\epsfig{figure=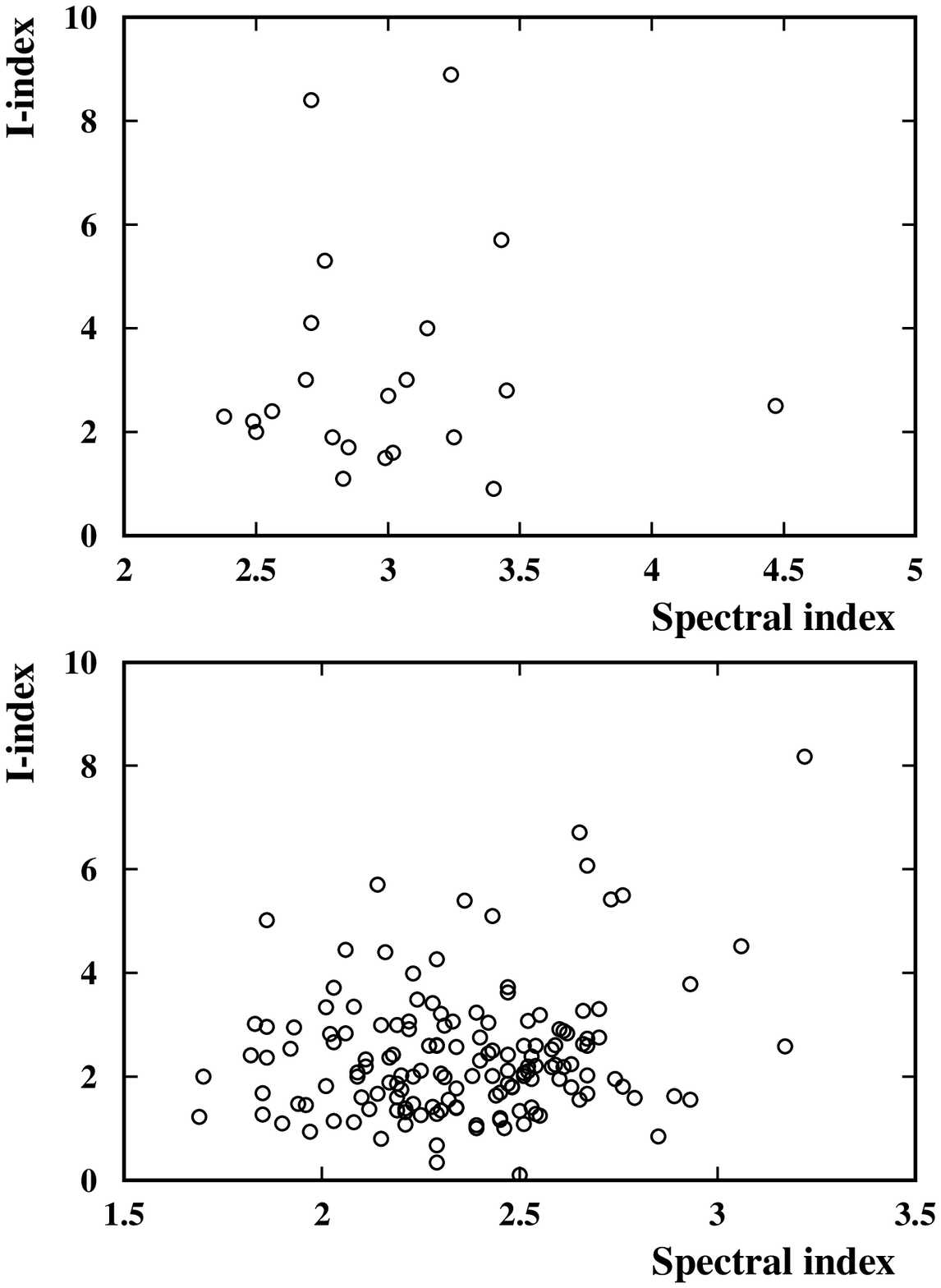,width=8.8cm,clip=}
\caption{
Correlation plots of spectral photon index versus variability I-index 
for our sample of 22 sources (upper panel) and the rest of 
151 unidentified and tentatively identified EGRET sources (lower panel).
The lower frame indicates a trend for higher variability with softer
spectra. 
} 
\label{fig:spec_Iindex}
\end{figure}

\section{Discussion}
By analyzing the COMPTEL data of unidentified and tentatively identified 
EGRET sources, and comparing their COMPTEL spectra with those from EGRET, 
we found a subgroup of 22 sources which show a spectral turnover or break
at MeV energies. At least two of them, but most likely the majority, 
have an emission maximum at energies between 1-100 MeV.
Variability analyses show that half of the sources are variable
above 100~MeV.
For the other half no significant time variability could be proven. 

The properties of the variable sources are reminiscent to the
observational properties of EGRET-detected blazars.
They are generally variable 
and their broadband (radio to \gray ) spectra are characterized by two 
spectral maxima. The high-energy one is for many sources  
located at MeV energies, e.g. 3C~273 \citep{Lichti95},
PKS~0528+134 \citep{Collmar97, Mukherjee99},
3C~279 \citep{Hartman01}. For details on their high-energy 
emission see e.g. \citet{Collmar01} and references therein. 
Because the EGRET blazars are typically high-latitude sources, 
variable unidentified high-latitude EGRET sources are generally 
considered to be of extra-galactic blazar origin.
Due to EGRET's decreasing sensitivity to variability with decreasing flux
they can appear as non-variable sources at low latitudes \citep{McLaughlin96},
where, due to the strong diffuse galactic emission, 
a higher significance level for source detection is required.
Therefore some of the low-latitude unidentified EGRET sources
may also be blazars.
Two sources of our sample, 3EG~J1424+3734 and 3EG~J2034-3110,
are high-latitude ($|$b$| >$ 30$^{\circ}$) sources.
3EG~J2034-3110 is variable and 3EG J1424+3734 is dubiously variable.
Our finding of a 'blazar-like' MeV spectrum provides further evidence 
for the blazar nature of these 2 sources, especially for the case of 
3EG~J2034-3110, for which time-variability is observed. 
The lack of an obvious blazar candidate at these source positions
can be explained by the calculations of \citet{Torres02}. 
The unidentified $\gamma$-ray sources could be a distant
weak $\gamma$-ray emitting 
blazars whose emission is amplified by gravitational microlensing.
The magnification factor is energy dependent and a broken energy 
spectrum is predicted.
Although the individual source properties of our sample sources
are reminiscent of blazars, 
their distribution on the sky is not. While the blazars are mainly detected 
at high latitudes, our source group concentrates towards the inner galaxy. 
Therefore we conclude, that at least the majority of these sources are 
not of blazar origin. 

Some \gray\ pulsars are also showing such an MeV-turnover spectrum. 
One example is PSR B1509-58.
\citet{Kuiper99} showed that its maximal luminosity is reached 
between 10 and 30~MeV, followed by a strong softening of its 
spectrum towards the EGRET band. 
In fact, PSR B1509-58 has not been detected by EGRET above 100 MeV,
while it is a strong COMPTEL source.
PSR~1509-58 is a young pulsar and its surface magnetic field can be 
inferred as at least 
3.1$\times$10$^{13}$ G (Kuiper et al. 1999). 
In such sources, electrons are accelerated up to relativistic 
energies, and subsequently produce \grays\ via curvature radiation 
and inverse-Comptonization of soft photons. 
In the vicinity of the magnetic pole, $\gamma$-rays
can be absorbed by the strong magnetic field via the photon splitting
process.
The latter process happens if the magnetic field is larger
than a critical value B $>$ 0.3~B$_{\rm{cr}}$,
 where B$_{\rm{cr}}$=4.413$\times$10$^{13}$~G 
is the surface magnetic field \citep{Harding97}. 
Its onset
has no energy threshold. For such a strong magnetic field, the polar
cap scenario predicts that photon splitting can become
 the dominant attenuation
process, resulting in a softening of the energy spectrum below
100 MeV, i.e. providing the observed MeV cutoff. It should be mentioned
that also in the competing outer gap
scenario such a spectral break at low $\gamma$-ray energies can be
accounted for.  Namely,
 Zhang and Cheng (2000b) applied their three dimensional outer
magnetosphere model to PSR B1509-58. They could reproduce 
the measured broad pulse profile and the measured
pulsed energy spectrum from the optical range up to \grays.  
This spectrum shows a power-law shape with a spectral bend above 1 MeV.
In the same work, Zhang and Cheng considered also
the case of the young Crab-like LMC pulsar PSR B0540-69,
which is also not detected by EGRET but below 10 keV stronger
than PSR B1509-58. The overall
characteristics of these two young pulsars appear rather similar.
By analyzing RXTE data, De Plaa et al. (2003) showed that PSR B0540-69
and the Crab pulsar have very similar spectral shapes up to about 50 keV,
where the Crab pulsar spectrum reaches its maximum luminosity. From these hard
X-rays to the EGRET energies above 100 MeV, the Crab spectrum softens
throughout the COMPTEL range (photon index $\sim$ -2.4; Kuiper et al. 2001).
De Plaa et al. discussed the spectral shapes of these 
three very young pulsars
($\leq$ 1.6 x 10$^3$ yr) which appear to be different from those of older
$\gamma$-ray pulsars like Vela and Geminga. The youngest pulsars are
strong(er) in the X-ray domain, but weak(er) above 100 MeV. Such a
spectral behaviour makes them less likely counterparts to the
unidentified EGRET sources,
 because their L$_{\rm X}$/L$_\gamma$ ratio is too high.
On the other
hand, all older established $\gamma$-ray pulsars like Vela, 
PSR~B1706-44, PSR~B1951+32, Geminga and PSR~B1055-52 
are weak X-ray emitters but have harder $\gamma$-ray spectra extending
up to the GeV range. The latter might be a selection effect. For
example, the weakly detected (above 50 MeV) $\gamma$-ray
pulsar PSR B0656+14 exhibits a steep spectrum (index $=$ -2.8$\pm$0.3)
at energies above 100~MeV \citep{Ramanamurthy96}. This pulsar has
a modest surface magnetic field of 4.7
x 10$^{12}$ G and a characteristic age of 1.1 x 10$^5$ years. It has not been
detected by COMPTEL below 30 MeV, implying a turnover of the soft EGRET
spectrum before the COMPTEL window, just like our sample.

Another pulsar type, the old, recycled weakly magnetized millisecond
pulsars, can also show such an MeV-turnover spectrum. Namely, a 4.9$\sigma$
detection has been claimed of the 2.3 ms pulsar
PSR J0218+4232 (Kuiper et al., 2000; 2002) at EGRET
energies. Its spin down parameters
give a rather old characteristic age of 4.6 x 10$^8$ years.
Its magnetic field of 4.3$\times$10$^8$~G is much weaker than 
that of standard $\gamma$-ray pulsars (e.g. Crab, Geminga).
A weak, but very hard X-ray spectrum was measured (photon index below 10 keV 
$>$ -1; Mineo et al. 2000) and a rather soft $\gamma$-ray spectrum
(photon index $\sim$ -2.6) was detected for EGRET above 100 MeV.
Current polar cap and outer gap models can account for the production
 of such high-energy radiation in millisecond pulsars, but have
difficulties in reproducing the observed spectral shape. Nevertheless,
the observed spectrum has the overall spectral shape of our source
sample, making millisecond pulsars viable candidate counterparts,
but only for the stable EGRET sources.
Millisecond pulsars are expected to be stable high-energy emitters
just like the normal radio pulsars, and contrary to case of 
accreting pulsars for which variability is naturally expected.

X-ray binaries (XRBs) have also been suggested as counterparts of
unidentified EGRET sources.
One XRB, the neutron star system Cen~X-3, has been detected
 as \gray\ emitter
during an activity period in 1994 (Vestrand et al. 1997). 
EGRET found a temporary 5$\sigma$ source which was 
positionally coincident with Cen~X-3. The 4.8s modulation of the
\grays, coinciding with the 4.8s rotation period of the neutron star, 
provided compelling evidence for the identification (Vestrand et al. 1997).
A hard power-law spectrum (photon index -1.81$\pm$0.37)
was measured  between 70~MeV and 10 GeV. The authors suggest that 
galactic X-ray binary systems may constitute a class
of highly variable GeV $\gamma$-ray sources.  
It is assumed that in Cen~X-3 we see the unabsorbed \gray\ spectrum. 
If in such a system the radiation region is surrounded by a condensed
soft photon field, e.g. in an accreting XRB, an energy-dependent absorption
of the $\gamma$-rays will occur resulting in a soft \gray\ spectrum. 
This scenario was proposed by \citet{Romero01} for the possible 
association of the EGRET source 
3EG J0542+2610 and the Be/X-ray transient A0535+26.
Such an absorption process was also studied by \citet{Wu93} for the 
XRB Cyg~X-3. Their simulations showed that the 
$\gamma$-ray spectrum can change
significantly when passing through the ambient soft X-ray 
field of an accreting
source. The 100~MeV -- 1~GeV emission will be absorbed. For both cases a 
soft $\gamma$-ray spectrum would be observable by EGRET,
if the absorption is not too strong. 

The emission processes of microquasars/blazars and extragalactic blazars
are -- in principle -- the same, however, on different time, space, and energy 
scales. Therefore one expects microquasars as potential counterparts 
of the unidentified EGRET sources. \citet{Paredes00} suggest the
microquasar LS~5039 to be the counterpart of 3EG~J1824-1514.
Calculations of microquasar spectra show that, depending on the strength 
of the jet Lorentz factor and magnetic field, microquasars could be
detected by EGRET, and that they could have their spectrum turnover 
at MeV energies (e.g. Kaufman Bernad$\acute{o}$ et al. 2002).      
However, no microquasar was definitely identified yet as an EGRET source. 
Given the current knowledge, the properties of some of our sources, 
spectral turnover and variability, could be matched by microquasars.

Binaries composed of early-type stars, like Wolf Rayet, Of and Be stars,
which produce strong stellar winds are proposed by 
Benaglia $\&$ Romero (2003) to be potential $\gamma$-ray emitters.
They argued that the electrons could be
accelerated to relativistic energies by the shocks generated in the colliding
wind region of the early-type binaries, and then cool via the process of
inverse-Compton scattering off the local soft photon field.
A low-energy cut off
of the electron spectrum or incomplete cooling of the electron 
population could lead to a turn over in the 
$\gamma$-ray spectrum at lower \gray\ energies.    
The early-type binaries are concentrated in the inner spiral arms 
of the Galaxy and variability is expected due to the changing geometry. 
However, according to the investigation of Romero et al. (1999) on the
 spatial correlation of low-latitude unidentified EGRET sources
with early-type stars,  
none of  the 22 sources of our sample is
spatially coincident with any early-type binary. 

Another class of sources proposed to be candidates for unidentified EGRET 
sources are supernova remnants (SNRs) (e.g. Esposito et al. 1996). In SNRs
electrons and/or protons can be accelerated by the Fermi process
to relativistic energies or appear in the outflows of a pulsar,
if one is embedded in the SNR. Subsequently $\gamma$-rays can be produced via
processes of inverse-Compton scattering, relativistic bremsstrahlung,
synchrotron emission, or $\pi_0$ decay.
Since the first-order Fermi process has a monotonic evolution of energies
with time and the acceleration time for the maximum energy is
limited by the age of
the remnant, it is generally thought that SNRs should be  
stable $\gamma$-ray sources, at least on a time scale of several years.
SNRs are spatially large, have ambient surrounding matter and 
a relatively weak magnetic field. A spectral turnover at 
MeV energies as found for 
our sample could be generated by inverse-Comptonization of soft photons,
by having the maximum of the synchrotron emission at MeV energies
(e.g. Crab nebula),
or by the decay of $\pi_{0}$ particles (generated in p-p collisions).
The latter one
would result in a broad spectral bump centered around 68 MeV.
Four sources of our sample, 3EG~J1639-4702, 1823-1314, 1837-0423, 1903+0550, 
are positionally coincident with SNRs.
3EG~J1823-1314 and 3EG~J1837-0423 are variable (Table~1) and therefore 
are unlikely counterparts of the SNRs. 
For 3EG~J1639-4702 and 3EG~J1903+0550 no variability is proven,
therefore they remain potential counterparts for the EGRET \gray\ sources.
Their possible associations with SNRs and coinciding radio pulsars 
have been discussed by  Torres et al. (2003).
With respect to radio pulsars, they concluded based on the energetics,
that for 3EG~J1639-4702 only one of the coinciding radio pulsar,
 PSR~J1637-4642,
has the potential to be the counterpart.
For 3EG~J1903+055 they did not find a potential pulsar counterpart, although 
they note that one of the coinciding pulsar lacks information and therefore
can not be judged.  
With respect to \gray\ emission due to $\pi_{0}$-decay, Torres et al. point out
that the SNRs are too far away to generate the observed flux. Even considering
an enhancement of the \gray\ production due to interactions of the 
accelerated nuclei with nearby dense molecular clouds, they find the 
identifications to be unlikely.  
These two unidentfied sources could be associated with a pulsar wind nebula, 
analogous to the case of the Crab for which the electrons might be
accelerated in the inner nebula and subsequently are generating a
synchrotron spectrum
which cuts off at MeV energies (de Jager et al.,1996). 
Their fluxes above 100~MeV have about the level of the Crab 
nebula. Because with distances of at least 8~kpc
(Torres et al., 2003), i.e. at least 4 times the distance to the Crab, they 
would have to be significantly more powerful at EGRET energies 
than the Crab nebula. 
However, this could be possible, if their synchrotron cut off would be shifted 
to higher energies (higher characteristic synchrotron energy) compared to 
the Crab nebula. 

Because for many unidentified EGRET sources no obvious candidate counterpart
exists, \citet{Romero99} mentioned isolated rotating
black holes, standard (Kerr) or charged (Kerr-Newman) ones,
as possible counterparts.
They could accrete from the diffuse interstellar medium (ISM).
Changes in the density of the
ISM would result in a variable \gray\ flux.
One such source, 3EG~J1828+0142, was modeled by \citet{Punsly00} 
by assuming an isolated Kerr-Newman black hole origin.
The model predicts a steep synchrotron self-Compton spectrum
with a spectral maximum at MeV energies, i.e. matching the spectral properties 
of our source sample. 
Including 3EG~J1828+0142, seven of our sample sources (3EG~J0520+2556,
0546+3948, 1638-5515, 1735-1500, 1741-2312, 1800-0146, 1828+0142)
are located at low galactic latitudes 
($|$b$|$$<$ 10$^{\circ}$) and lack any positional coincidence with 
galactic objects of potential counterpart nature.
Two of them, 3EG~J1735-1500 and 3EG~1828+0142, are significantly variable
and therefore match the anticipated \gray\ properties of 
these exotic objects.

\section{Conclusion}
By analyzing the contemporary COMPTEL 
observations of all unidentified  or only tentatively identified 
EGRET sources, we found a subgroup of 22 sources for which 
we can provide spectral constraints for source modelling.
Their spectra have to turn over between $\sim$1 MeV and 100
MeV, and at least two of them, but most likely the majority,
have their maximum luminosities somewhere in this energy band.
Most of the sources are not detected by COMPTEL, however the simultaneously 
derived upper limits require the spectral bending. 
These sources have rather steep energy spectra in the EGRET band, 
and seem to be preferentially located in the inner galaxy, 
especially at low latitudes ($|$b$|$ $<$ 30$^{\circ}$).
Variability studies reveal that half of them are significantly variable
above 100~MeV. 
Potential counterparts have to conform to these observational results.
A blazar origin for the two high-latitude sources in this
sample seems to be likely. 
Viable candidate counterparts for the steady low-latitude sources are:
1) young (age $<$ 10$^6$ years) pulsars, although the youngest with the
strongest magnetic fields might be too strong in the X-ray domain, 2) 
old, recycled millisecond pulsars with a weak magnetic field (like PSR
J0218+4232), and 3) SNRs and pulsar wind nebula whose synchrotron spectra
are peaking at MeV energies (like the Crab nebula). 
For the variable low-latitude sources, XRBs, in particular 
microquasars/blazars by assuming a spectral analogy to
the extragalactic objects, and isolated BHs would match 
 the requirements.  
Case by case studies might reveal further insights in the nature
of individual sources.

\begin{acknowledgements}
This research was supported by the German government through DLR grant 
50 QV 9096 8, by NASA under contract NAS5-26645, and by the Netherlands
Organization for Scientific Research NWO. 
S. Zhang was also subsidized by the 
Special Funds for Major State Basic Research Projects
and by the National Natural Science Foundation of China
under grant 10373013. 

\end{acknowledgements}


\end{document}